# Creek: Low-latency, Mixed-Consistency Transactional Replication Scheme

Tadeusz Kobus, Maciej Kokociński, and Paweł T. Wojciechowski
Institute of Computing Science
Poznań University of Technology
60-965 Poznań, Poland
{Tadeusz.Kobus,Maciej.Kokocinski,Pawel.T.Wojciechowski}@cs.put.edu.pl

**Abstract**

In this paper we introduce *Creek*, a low-latency, eventually consistent replication scheme that also enables execution of strongly consistent operations (akin to ACID transactions). Operations can have arbitrary complex (but deterministic) semantics. Similarly to state machine replication (SMR), Creek totally-orders all operations, but does so using two different broadcast mechanisms: a timestamp-based one and our novel *conditional atomic broadcast (CAB)*. The former is used to establish a tentative order of all operations for speculative execution, and it can tolerate network partitions. On the other hand, CAB is only used to ensure linearizable execution of the strongly consistent operations, whenever distributed consensus can be solved. The execution of strongly consistent operations also *stabilizes* the execution order of the causally related weakly consistent operations. Creek uses multiversion concurrency control to efficiently handle operations' rollbacks and reexecutions resulting from the mismatch between the tentative and the final execution orders. In the TPC-C benchmark, Creek offers up to 2.5 times lower latency in returning client responses compared to the state-of-the-art speculative SMR scheme, while maintaining high accuracy of the speculative execution (92-100%).

## 1 Introduction

A lot of research has been devoted in the last years to eventually consistent replicated data stores, such as modern NoSQL databases (e.g., [1] [2] [3]). These systems are scalable, guarantee high availability, and provide low response times, unlike their strongly consistent counterparts (e.g., traditional relational database systems [4]), Ensuring low latency in serving client requests is especially important for today's globally accessible services running on the Internet, even at the cost of occasionally providing clients with partially incomplete or inconsistent responses [5] [6] [7]. Moreover, in existing NoSQL databases, the low latency and the ability of a system to scale to hundreds or thousands of nodes comes at the cost of reduced flexibility in terms of offered semantics (e.g., they support only CRUD operations). In effect, such systems do not suit all applications, and the process of developing services on top of such systems is much more costly than using traditional database systems.[1]

Programmers who build services on top of eventually consistent data stores, must anticipate the possibility of working on skewed data, which means properly handling all edge cases. Failing to do so results in unexpected errors which are often difficult and costly to fix later in the development process. Moreover, the limited semantics of such data stores means that programmers have to implement custom mechanisms for synchronizing concurrent accesses to shared data for situations in which the *act now, apologize later* pattern cannot be used (e.g., a payment operation typically should end with a clear indication of whether the operation succeeded or failed). This is why programmers, who are used to traditional relational database systems, miss the fully fledged support for serializable transactions, which, naturally, cannot be provided in a highly-available fashion [9] [10]. Therefore, in recent years, various NoSQL vendors started adding (quasi) transactional support to their systems. These add-on mechanisms are often very prohibitive and do not perform well. For example, both Riak and Apache Cassandra do not offer cross-object/cross-record transactions [11] [12]. Additionally, Riak allows strongly consistent (serializable) operations to be performed only on distinct data [11], whereas using the *light weight transactions* in Apache Cassandra on data that are accessed at the same time in the regular, eventually consistent fashion leads to undefined behaviour [13].

In this paper we present *Creek*, a novel fault-tolerant replication scheme designed from scratch to seamlessly incorporate linearizable transactions into an eventually consistent environment for improved performance and semantics. More precisely, Creek's main features are as follows:

1. Creek supports execution of client submitted *operations* of arbitrary semantics, e.g., arbitrarily complex, deterministic SQL transactions.
2. An operation can be executed either in a strongly consistent fashion (akin to ACID transactions), or in a way

---

[1]Using conflict-free replicated data types (CRDTs) [8] may help to some degree, as CRDTs offer clear guarantees and hassle-free state convergence. However, CRDTs have limited semantics: they either require that the operations they define always commute, or that there exist some commutative, associative and idempotent state merge procedures.

that trades consistency for low latency, as in eventually consistent systems; we call these operations *strong* and *weak*, respectively.
3. Strong and weak operations can operate on the same data at the same time.
4. Each Creek replica totally orders all operations it knows about and speculatively executes them to achieve high throughput and reduce latency. Additionally, for a strong operation *op* we define:
    - a *tentative* response–yielded by the first speculative execution of *op* (on the replica that received *op* from the client),
    - a *stable* response–obtained during the execution of *op* after the inter-replica synchronization or during a speculative execution of *op*, which is consistent with the final execution order of *op* (as established during the inter-replica synchronization).
5. The execution of strong and weak operations is causally binded, so that the execution of operations of different types is not entirely independent. More precisely, for any strong operation *op*, its stable response will reflect the execution of at least all weak operations that would be visible to *op*'s speculative execution if it happened as soon as possible (at the time when *op* is initially received by a system replica from the client; the execution of *op* could be deferred, when the replica has a backlog of unprocessed operations).

Creek executes each weak operation speculatively (thus ensuring a low-latency response), in the order corresponding to the operation's timestamp that is assigned once a Creek replica receives an operation from a client. A strong operation is also executed speculatively (yielding a *tentative response*), but for such an operation Creek also runs a round of inter-replica synchronization to determine the final operation execution order. Afterwards the strong operation's *stable* response can be returned to the client. The execution of a strong operation also *stabilizes* the execution order of the causally related weak operations.

The final operation execution order is established using our new total order protocol, *conditional atomic broadcast (CAB)*, which is based on indirect consensus [14]. The messages broadcast using CAB are as small as possible and are limited to the identifiers of strong operations. The contents of all (weak and strong) operations are disseminated among replicas using only a gossip protocol (reliable broadcast). Creek leverages multiversioning scheme [15] to facilitate concurrent execution of operations and to minimize the number of necessary operation rollbacks and reexecutions.

Creek can gracefully tolerate node failures, because CAB can be efficiently implemented by extending quorum-based protocols, such as Multi-Paxos [16]. When network partitions occur, replicas within each partition are still capable of executing weak operations and obtaining tentative responses to strong operations, and converging to the same state, once the stream of client requests ceases. Formally Creek guarantees linearizability [17] for strong operations, and fluctuating eventual consistency [18] for weak operations.

We use the TPC-C benchmark [19] to assess the performance of Creek. The TPC-C's *Payment* transactions are executed as strong operations because, intuitively, they should provide the client with a definite (stable) response. All other transactions are executed as weak operations. We compare Creek's performance to the performance of other replication schemes that enable arbitrary transactional semantics and from which Creek draws inspiration: Bayou [20], SMR [21] [22], and a state-of-the-art speculative SMR scheme [23]. By leveraging the multicore architecture of modern hardware, Creek easily outperforms SMR and Bayou. Creek provides throughput that is on-par with the speculative SMR scheme, but exhibits much lower latencies in serving client requests (up to 2.5 times lower for weak transactions and up to 20% lower for strong transactions). In the vast majority of cases (92-100%, depending on the scenario), the effects of the speculative request execution correspond to the final execution order as established by solving distributed consensus among Creek replicas. Note that similarly to the schemes we test Creek against, Creek also assumes full replication of data. In the paper we outline the design of a partially replicated version of Creek that we plan to implement in the future.

The reminder of the paper is structured as follows. We discuss related work in Section 2. Then, in Section 3 we specify the system model and CAB, our new broadcast protocol. In Section 4 we present the Creek scheme in detail and in Section 5 show how CAB can be efficiently implemented. Finally, we experimentally evaluate Creek in Section 6 and conclude in Section 7.

## 2 Related work

In the past several researchers attempted to incorporate transactional semantics into eventually consistent systems. The proposed solutions usually assumed weaker guarantees for transactional execution (e.g., [24] [25] [26]), or restricted the semantics of transactions (e.g., [27] [28]). Interestingly, the first eventually consistent transactional systems, i.e., *Bayou* [20] and the implementations of *Eventually-Serializable Data Services (ESDS)* [29], followed a different approach. In these systems, each server speculatively total-orders all received client requests without prior agreement with other servers. The final request serialization is established by a primary server. In case the speculation is wrong, some of the requests are rolled back and reexecuted (Bayou), or, to obtain the response for a new client request, much of the requests whose final execution order is not yet established are repeatedly reexecuted (ESDS). Understandably, such implementations cannot perform and scale well. Moreover, they are not fault-tolerant because of the reliance on the primary. However, these systems have one very important trait: reasoning about their behaviour is relatively easy and comes intuitively, because, similarly to *state machine replication (SMR)* [21] [22], which executes all requests on all servers in the same order, on each server there is always a single serialization of all client requests the server knows about. In this sense, Bayou, ESDS and SMR serve as direct inspirations for Creek.

There are numerous subtle characteristics that make Creek a more versatile system. Unlike Creek, in Bayou updating transactions do not provide return values. Bayou features *dependency check* and *merge procedure* mechanisms, that allow the system to perform application-level conflict detection and resolution. In Creek, we do not make any (strong) assumptions on the semantics of operations handled by the system (see also Section 4.2), but these mechanisms can be emulated on the level of operation specification, if required.

ESDS allows a client to attach to an operation an arbitrary causal context that must be satisfied before the operation is executed. Creek can be easily extended to accommodate the full specification of ESDS. Interestingly, the basic implementation of ESDS [29] does not incrementally evaluate the service state with each operation execution. Instead, in order to obtain a response to an operation, ESDS first creates a provisional state by reexecuting (some of) the previously submitted operations. Local computation is assumed to happen instantaneously. Naturally, this assumption is not realistic, so an optimized version of ESDS has been implemented, which limits the number of costly operation reexecutions and saves network usage [30].

As we mentioned above, there are some similarities between Creek and SMR. In basic SMR, every replica sequentially executes all client requests (transactions) in the same order [21] [22]. For this, SMR usually relies on the *atomic broadcast (AB)* (also called *total order broadcast*) protocol to ensure that all servers deliver the same set of requests in the same order. The speculative schemes based on SMR (e.g., [31] [32] [23]) start the execution of a request before the final request execution order is established, to minimize latency in providing response to the client. However, even though for some requests the system might produce responses before the final requests execution order is established, the responses are withheld until the system ensures that the execution is serializable. Hence, these approaches do not guarantee low-latency responses.

To enable SMR to scale, some schemes (e.g., [33] [34]) utilize *partial replication*, in which data is divided into *partitions*, each of which can be accessed and modified independently. In Section 4.5 we discuss how Creek can be extended to support partial replication.

In Section 6 we compare the performance of Creek to the performance of Bayou, SMR as well as Archie [23], the state-of-the-art speculative SMR scheme. Archie uses a variant of optimistic atomic broadcast to disseminate requests among servers that guarantees that in stable conditions (when the leader of the broadcast protocol does not change), the optimistic message delivery order is the same as the final one. Similarly to Creek, Archie utilizes multiversioning scheme and takes full advantage of the multi-core hardware.

We are aware of several systems that similarly to Creek feature requests that can be executed with different consistency guarantees. The system in [35] enables enforcing two kinds of stronger guarantees than causal consistency, by either a consistent numbering of requests, or the use of the three-phase-commit protocol. Unlike Creek, the system does not enforce a single total order of all client requests. Zeno [36] is very similar to Bayou, but it has been designed to tolerate Byzantine failures. Li *et al.* [37] demonstrate Gemini, a replicated system that satisfies *RedBlue consistency*. Gemini ensures causal consistency for all operations, but unlike the strong (red) operations, the weak (blue) operations must commute with all other operations. Burckhardt *et al.* [38] describe *global sequence protocol (GSP)*, in which client applications perform operations locally and periodically synchronize with the cloud, the single source of truth. The cloud is responsible for establishing the final operation execution order. Changes to the execution order might lead to operation rollbacks and reexecutions in the client applications. When the cloud is unavailable, GSP does not guarantee progress: the clients can issue new operations that are executed locally, but they are not propagated to other clients. In effect, when the cloud is down, each client is forced to work in a separate network partition.

Recently there have been attempts to formalize the guarantees provided by Bayou and systems similar to it. Shapiro *et al.* [39] propose a (rather informal) definition of *capricious total order*, in which each server total-orders all operations it received, without prior agreement with others. In [40], Girault *et al.* propose a more formal property called *monotonic prefix consistency*. The definition is, however, limited to systems that, unlike Creek, only feature read-only operations and updating operations that do not have a return value. To formalize Creek's correctness we use the framework and a property called *fluctuating eventual consistency* that were introduced in [18] (see Section 4.3).

## 3 Model

### 3.1 Processes and crashes

We consider a fully asynchronous, message-passing system consisting of a set $\Pi = \{p_1, ..., p_n\}$ of processes (also called *replicas*), to which external clients submit requests in the form of operations (also called *transactions*) to be executed by the processes. We model each process as a state automaton, that has a local state and, in reaction to events, executes steps that atomically transition the replica from one state to another. We consider a crash-stop failure model, in which a process can crash by ceasing communication. A replica that never crashes is said to be *correct*, otherwise it is *faulty*.

### 3.2 Reliable broadcast

Replicas communicate via reliable channels. Replicas can use *reliable broadcast (RB)* [41], that is defined through two primitives: RB-cast and RB-deliver. Intuitively, RB guarantees that even in case a faulty replica RB-casts some message $m$ and it is RB-delivered by at least one correct replica, all other correct replicas eventually RB-deliver $m$. Formally, RB requires: (1) *validity*–if a correct replica RB-casts some message $m$, then the replica eventually RB-delivers $m$, (2) *uniform integrity*–for any message $m$, every process RB-delivers $m$ at most once and only if $m$ was previously RB-cast, and (3) *agreement*–if a correct replica RB-delivers some message $m$, then eventually all replicas RB-deliver $m$.

### 3.3 *Conditional atomic broadcast*

Now we formally define CAB, our novel broadcast primitive that is used by Creek to handle strong operations. In Section 5 we discuss how CAB can be efficiently implemented.

Similarly to *atomic broadcast (AB)* (also called *total order broadcast*) [42], CAB enables dissemination of messages among processes with the guarantee that each process delivers all messages in the same order. Unlike AB, however, CAB allows a process to *defer* the delivery of a message until a certain condition is met (e.g., certain related network communication is completed). To this end, CAB defines two primitives: CAB-cast($m, q$), which is used by processes to broadcast a message $m$ with a test predicate $q$ (or simply, predicate $q$), and CAB-deliver($m$) to deliver $m$ on each process but only when the predicate $q$ is satisfied. Since $q$ might depend on $m$, we write $q(m) = true$ if $q$ is evaluated to *true* (on some process $p_i$). $q$ must guarantee *eventual stable evaluation*, i.e., $q$ needs to eventually evaluate to *true* on every correct process and once $q(m) = true$ on replica $p_i$, then $q(m)$ never changes to *false* on $p_i$. Otherwise, not only CAB would not be able to terminate, but different processes could CAB-deliver different sets of messages. We formalize CAB through the following requirements:

- *validity*: if a correct process $p_i$ CAB-casts a message $m$ with predicate $q$, and eventual stable evaluation holds for $q$, then $p_i$ eventually CAB-delivers $m$,
- *uniform integrity*: for any message $m$ with predicate $q$, every process $p_i$ CAB-delivers $m$ at most once, and only if $(m, q)$ was previously CAB-cast and $q(m) = true$ at $p_i$,
- *uniform agreement*: if a process $p_i$ (correct or faulty) CAB-delivers $m$ (with predicate $q$), and eventual stable evaluation holds for $q$, then all correct processes eventually CAB-deliver $m$ (with $q$),
- *uniform total order*: if some process $p_i$ (correct or faulty) CAB-delivers $m$ (with predicate $q$) before $m'$ (with predicate $q'$), then every process $p_j$ CAB-delivers $m'$ (with $q'$) only after it has CAB-delivered $m$ (with $q$).

### 3.4 Weak and strong operations

As we have already outlined earlier, clients may issue two kinds of operations: *weak* and *strong*. *Weak* operations are meant to be executed in a way that minimizes the latency in providing a response to the client. Hence, we require that a replica that received a weak operation executes it immediately in an eventually consistent fashion on the local state and issues a response to the client without waiting for coordination with other replicas (using a gossip protocol). Other replicas also execute the operation in a eventually consistent way as soon as they receive the operation. This behaviour is necessary (but not sufficient) to ensure that in the presence of network partitions (when communication between subgroups of replicas is not possible for long enough), the replicas' states in each subgroup converge once the stream of operations submitted by clients ceases. Naturally, a response to a weak operation might not be correct, in the sense that it might not reflect the state of replicas once they synchronize. On the other hand, a replica returns to a client a (stable) response to a *strong* operation only after the replicas synchronize and achieve agreement on the final operation execution order (relative to other, previously handled operations). Achieving agreement among replicas requires solving distributed consensus. We assume availability of failure detector $\Omega$, the weakest failure detector capable of solving distributed consensus in the presence of failures [43].

## 4 Creek

### 4.1 Overview

Each Creek replica totally-orders all operations it knows about. In order to keep track of the total order, a replica maintains two lists of operations: *committed* and *tentative*. The *committed* list encompasses strong operations and the *stabilized* weak operations, i.e., operations that belonged to the causal context of some committed strong operation and thus whose final execution order has been established (see also below). The order of operations on the *committed* list is determined by the synchronization of replicas using CAB. Hence the order is the same across all replicas. The messages disseminated using CAB are as small as possible and only include the identifiers of the strong operations. On the other hand, the *tentative* list encompasses strong operations whose final execution order has not yet been determined (such strong operations are to be speculatively executed) and weak operations that are not yet stabilized. The operations on the *tentative* list are sorted using the operations' timestamps. A timestamp is assigned to an operation as soon as a Creek replica receives it from a client. Then the operation is disseminated by the replica among other replicas using a gossip protocol so that all replicas can independently execute it (each replica executes operations according to the order of operations on the replica's concatenated *committed* and *tentative* lists). For a strong operation *op*, the replica $p_i$ that receives it performs two additional steps. Firstly, $p_i$ CAB-casts the identifier of *op*. Secondly, to the message sent using the gossip protocol, $p_i$ also attaches the *causal context* of *op*, i.e., the current set of weak operations on the *tentative* list that have a lower timestamp than *op*. The causal context represents the weak operations that $p_i$ knows about, and that must be executed by each replica before the *op* is speculatively executed (hence, a replica might have to postpone the execution of *op*). Each replica will also stabilize these operations once the replica CAB-delivers a message with *op*'s identifier (as explained in Section 4.2).

A Creek replica continually executes one by one operations in the order determined by the concatenation of the two lists: *committed*·*tentative*. An operation $op \in committed$, once executed, will not be executed again as its final operation execution order is determined. It is not necessarily the case for operations in the *tentative* list. It is because a replica adds operations to the *tentative* list (rearranging it if necessary) as they are delivered by a gossip protocol. Hence, a replica might execute some $op \in tentative$, an then, in order to maintain the proper order on the *tentative* list, the replica might be forced to roll *op* back, execute a just received operation *op*′ (which has lower timestamp than *op*), and execute *op* again. However, if replica clocks are relatively closely synchronized and the delays in network communication do not deviate much,[2] Creek efficiently handles incoming operations. In the majority of cases, the operations received by the gossip protocol are received in an order corresponding to their timestamps. Hence the received operations are simply added to the tail of the *tentative* list and scheduled for execution. Moreover, the order of speculative execution of operations closely matches the final operation execution order as established by CAB. It means that typically no operation rollbacks or reexecutions are necessary: a strong operation *op* whose identifier has just been CAB-delivered, is simply *moved*, together with the weak operations that belong to *op*'s causal context, from the head of the *tentative* list to the tail of the *committed* list.

In the subsequent sections we present the basic version of our scheme in more detail (Section 4.2), argue about its correctness (Section 4.3), discuss how we optimized its performance (Section 4.4) and finally how we plan to extend it in the future to encompass partial replication (Section 4.5).

### 4.2 Basic scheme–detailed specification

Our specification of Creek, shown in Algorithm 1, is rooted in the specification of the Bayou protocol [20] presented in [18]. We assume that clients submit requests to the system in the form of *operations* with encoded arguments (line 15), and await responses. Operations are defined by a specification of a (deterministic) replicated data type $\mathcal{F}$ [44] (e.g., *read*/*write* operations on a register, list operations, such as *append*, *getFirst*, or arbitrary SQL queries/updates). Each operation is marked as weak or strong (through the *strongOp* argument). Operations are executed on the *state* object (line 4), which encapsulates the state of a copy of a replicated object implementing $\mathcal{F}$. Algorithm 2 shows a pseudocode of StateObject, a referential implementation of *state* for arbitrary $\mathcal{F}$ (a specialized one can be used for a specific $\mathcal{F}$). We assume that each operation can be specified as a composition of read and write operations on registers (objects) together with some local computation. The assumption is sensible, as the operations are executed locally, in a sequential manner, and thus no stronger primitives than registers (such as CAS, fetch-and-add, etc.) are necessary. The StateObject keeps an undo log which allows it to revoke the effects of any operation executed so far (the log can be truncated to include only the operations on the *tentative* list).

Upon invocation of an operation *op* (line 15), it is wrapped in a Req structure (line 17) that also contains the current timestamp (stored in the *timestamp* field) which will be used to order *op* among weak operations and strong operations executed in a tentative way, and its unique identifier (stored in the *id* field), which is a pair consisting of the Creek replica number $i$ and the value of the monotonically increasing local event counter *currEventNo*). Such a package is then distributed (gossiped) among replicas using a reliable broadcast protocol, line 23; we simply say that *op* has been RB-cast and later RB-delivered; in lines 22 and 24 we simulate immediate local RB-delivery of *op*). If *op* is a strong operation, we additionally attach to the message the *causal context* of *op*, i.e., the identifiers of all operations that have already been RB-delivered by the replica and which will be serialized before *op* (line 19).[3] This information can be effectively stored in a dotted version vector (dvv) [45], which is logically a set of pairs of a replica identifier and an event number (in the *causalCtx* variable, line 7, a replica maintains the identifiers of all operations the replica *knows about*, see the routines in lines 26 and 32). For a strong operation, the replica also CAB-casts the operation's identifier with a test predicate specified by the *checkDep* function (line 20). By specification of CAB, *checkDep*(*id*) (line 10) is evaluated by CAB on each replica at least two times: (1) when solving distributed consensus on a concrete operation identifier *id* that is about to be CAB-delivered, and then, (2) after the decision has been reached, in an attempt to CAB-deliver *id* locally (*checkDep*(*id*) is guaranteed to eventually evaluate to

---
[2]Creek's correctness does not depend on such assumptions.

[3]Operations serialized before *op* include all operations RB-delivered by the replica whose final operation execution order is already established, and other weak operations whose timestamp is smaller than *op*'s timestamp. Later we explain why the causal context of *op* cannot include the identifiers of any strong operations whose final execution order is not yet determined.

**Algorithm 1** The Creek protocol for replica $i$

```
 1: struct Req(timestamp : int, id : pair⟨int, int⟩,
              op : ops(F), strongOp : boolean, causalCtx : dvv)
 2: operator <(r : Req, r' : Req)
 3:   return (r.timestamp, r.id) < (r'.timestamp, r'.id)
 4: var state : StateObject
 5: var currEventNo : int
 6: var committed, tentative : list⟨Req⟩
 7: var causalCtx : dvv                         // logically set⟨pair⟨int, int⟩⟩
 8: var executed, toBeExecuted, toBeRolledBack : list⟨Req⟩
 9: var reqsAwaitingResp : map⟨Req, Resp⟩
10: function checkDep(id : ⟨int, int⟩)
11:   var r = x : (x ∈ (committed · tentative) ∧ x.id = id)
12:   if r = ⊥ then
13:     return false
14:   return r.causalCtx ⊆ causalCtx
15: upon invoke(op : ops(F), strongOp : boolean)
16:   currEventNo = currEventNo + 1
17:   var r = Req(currTime, (i, currEventNo), op, strongOp, ⊥)
18:   if strongOp then
19:     r.causalCtx = causalCtx \ {x.id|x ∈ tentative ∧ r < x}
20:     CAB-cast(r.id, checkDep)
21:   else
22:     causalCtx = causalCtx ∪ {r.id}
23:     RB-cast(r)
24:   adjustTentativeOrder(r)
25:   reqsAwaitingResp.put(r, ⊥)
26: upon RB-deliver(r : Req)
27:   if r.id.first = i then                    // r issued locally
28:     return
29:   if ¬r.strongOp then
30:     causalCtx = causalCtx ∪ {r.id}
31:   adjustTentativeOrder(r)
32: upon CAB-deliver(id : pair⟨int, int⟩)
33:   var r = x : (x ∈ tentative ∧ x.id = id)
34:   causalCtx = causalCtx ∪ {r.id}
35:   commit(r)
36: procedure adjustTentativeOrder(r : Req)
37:   var previous = [x|x ∈ tentative ∧ x < r]
38:   var subsequent = [x|x ∈ tentative ∧ r < x]
39:   tentative = previous · [r] · subsequent
40:   var newOrder = committed · tentative
41:   adjustExecution(newOrder)
42: procedure adjustExecution(newOrder : list⟨Req⟩)
43:   var inOrder = longestCommonPrefix(executed, newOrder)
44:   var outOfOrder = [x|x ∈ executed ∧ x ∉ inOrder]
45:   executed = inOrder
46:   toBeExecuted = [x|x ∈ newOrder ∧ x ∉ executed]
47:   toBeRolledBack = toBeRolledBack · reverse(outOfOrder)
48: procedure commit(r : Req)
49:   var committedExt = [x|x ∈ tentative ∧ x.id ∈ r.causalCtx]
50:   var newTentative = [x|x ∈ tentative ∧ x ∉ committedExt ∧ x ≠ r]
51:   committed = committed · committedExt · [r]
52:   tentative = newTentative
53:   var newOrder = committed · tentative
54:   adjustExecution(newOrder)
55:   if reqsAwaitingResp.contains(r) ∧ r ∈ executed then
56:     return reqsAwaitingResp.get(r) to client (as stable response)
57:     reqsAwaitingResp.remove(r)
58: upon toBeRolledBack ≠ []
59:   var [head] · tail = toBeRolledBack
60:   state.rollback(head)
61:   toBeRolledBack = tail
62: upon toBeRolledBack = [] ∧ toBeExecuted ≠ []
63:   var [head] · tail = toBeExecuted
64:   var response = state.execute(head)
65:   if reqsAwaitingResp.contains(head) then
66:     if ¬head.strongOp then
67:       return response to client
68:       reqsAwaitingResp.remove(head)
69:     else if head.strongOp ∧ head ∈ tentative then
70:       reqsAwaitingResp.put(head, response)
71:       return response to client (as tentative response)
72:     else                          // head.strongOp ∧ head ∈ committed
73:       return response to client (as stable response)
74:       reqsAwaitingResp.remove(head)
75:   executed = executed · [head]
76:   toBeExecuted = tail
```

**Algorithm 2** StateObject for some replica

```
 1: var db : map⟨Id, Value⟩
 2: var undoLog : map⟨Req, map⟨Id, Value⟩⟩
 3: function execute(r : Req)
 4:   var undoMap : map⟨Id, Value⟩
 5:   execute r.op line by line
 6:   upon read(id : Id)
 7:     return db[id]
 8:   upon write(id : Id, v : Value)
 9:     if undoMap[id] = ⊥ then
10:       undoMap[id] = db[id]
11:     db[id] = v
12:   upon return(response : Resp)
13:     undoLog[r] = undoMap
14:     return response
15: function rollback(r : Req)
16:   var undoMap = undoLog[r]
17:   for (id, v) ∈ undoMap do
18:     db[id] = v
19:   undoLog = undoLog \ (r, undoMap)
```

*true* on every correct replica). The function checks whether the replica has already RB-delivered the strong operation *op* identified by *id*, and if so, whether it has also already RB-delivered all operations that are in the causal context of *op*. Note that a replica will CAB-deliver *op*'s identifier only if it had already RB-delivered *op*'s Req structure.

When an operation *op* is RB-delivered (line 26), the replica adds its identifier to *causalCtx* if *op* is a weak operation (line 30), and then uses *op*'s *timestamp* to correctly order *op* among other weak operations and strong operations targeted for speculative execution (on the *tentative* list of Reqs, line 39). Then the new operation execution order is established by concatenating the *committed* list and the *tentative* list (line 40; recall that the *committed* list maintains the Req structures for all operations, whose final execution order has already been established). Then, the adjustExecution function (line 42) compares the newly established operation execution order with the order in which some operations have already been executed (see the *executed* variable). Operations, for which the orders are different, are rolled back (in the order opposite to their execution order) and reexecuted. In an ideal case, *op* is simply added to the end of the *toBeExecuted* list, and awaits execution.[4] To limit the number of possible rollbacks, local clocks (used to generate timestamps for Req structures) should not deviate too much from each other.

When an operation *op*'s identifier is CAB-delivered (line 32), the replica can *commit op*, i.e., establish its final execution order. To this end, the replica firstly *stabilizes* some of the operations, i.e., moves the Req structures of all operations included in the causal context of *op* from the *tentative* list to the end of the *committed* list (line 49). Then the replica adds *op*'s Req structure to the end of the *committed* list as well (line 51). Note that this procedure preserves the relative order in which weak operations from the causal context of *op* appear on the *tentative* list (the causal precedence of these operations in relation to *op* does not change). All operations not included in the causal context of *op* stay on the *tentative* list (line 50). As before, the adjustExecution function is called to mark some of the executed operations for rollback and reexecution (line 54). Note that in an ideal case, operations (including *op*) can be moved from the *tentative* to the *committed* list without causing any rollbacks or reexecutions. Unfortunately, if any (weak) operation submitted to some other replica is ordered in-between operations from the causal context of *op*, and some of these operations are already executed, rollbacks cannot be avoided in the basic version of Creek. In Section 4.4 we discuss how this situation can be mitigated to some extent.

---
[4]No rollbacks are also required when execution lags behind the RB-delivery of operations. Then the tail of the *toBeExecuted* list will undergo reordering.

Recall that the causal context of a strong operation *op* does not include the identifiers of any strong operations that are not yet committed. We cannot include such dependencies because, ultimately, the order of strong operations is established by CAB, which is unaware of the semantics and the possible causal dependency between messages sent through CAB. Hence, the order of strong operations established by CAB might be different from the order following from the causal dependency that we would have had defined. In principle, such dependencies could be enforced using Zab [46] or executive order broadcast [47]. However, these schemes would have to be extended to accommodate the capabilities of CAB. In Creek, the identifier *id* of a strong operation *op* is added to the global variable *causalCtx* (which we use to create a causal context for all strong operations) only upon CAB-deliver. But then we commit *op*, thus establishing its final execution order.

Operation rollbacks and executions happen within transitions specified in lines 58–61 and 62–76. Whenever an operation is executed on a given replica, the replica checks if the operation has been originally submitted to this replica (line 65). If so, the replica returns the (tentative or stable) response to the client. Note that in our pseudocode, before a client receives a stable response to a strong operation, it may receive multiple tentative responses, one for each time the operation is (re)executed. Sometimes the replica returns a stable response in the commit function (line 48). This happens when a strong operation has been speculatively executed in an order which is equivalent to its final execution order.

### 4.3 Correctness

In order to precisely capture the guarantees provided by Creek, we resort to the formal framework from [18], which the authors use to analyze the behaviour and then formalize the guarantees of the seminal Bayou protocol [20]. Creek's principle of operation is similar to Bayou's, so Creek also exhibits some of Bayou's quirky behaviour. Most crucially, Creek allows for *temporary operation reordering*, which means that the replicas may temporarily disagree on the relative order in which the operations submitted to the system were executed. In consequence, clients may observe operation return values which do not correspond to any operation execution order that can be produced by traditional relational database systems or typical NoSQL systems. As the authors prove, this characteristics is unavoidable in systems that mix weak and strong consistency. The basic version of Creek is also not free of the other two traits of Bayou mentioned in [18], namely *circular causality* and *unbounded wait-free execution* of operations. The former can be mitigated in a similar fashion as in Bayou.

Formally, the guarantees provided by Creek can be expressed using *Fluctuating Eventual Consistency (FEC)* [18], a property that precisely captures temporary operation reordering and is not tied to a concrete data type.[5] Below we argue why Creek satisfies FEC for weak operations and linearizability (LIN) [17] for strong operations. We use $\mathcal{F}$, a specification of a replicated data type, as a way to capture the semantics of the system. We consider *stable runs* of a replicated system, which means that solving consensus is possible (otherwise, we could only prove that Creek satisfies FEC for weak operations). The FEC and LIN properties target the set of weak and strong operations, respectively.

**Theorem 1.** *Creek satisfies FEC(weak, $\mathcal{F}$) $\wedge$ LIN(strong, $\mathcal{F}$) in stable runs for arbitrary $\mathcal{F}$.*

*Proof sketch.* Informally, in order to prove that Creek satisfies linearizability for strong operations, we need to show that the stable return values produced by Creek are such that they can be obtained by a sequential execution of all operations in some total order *S* that (1) respects the program order of every replica from the point of view of each strong operation, and (2) also respects the real-time order of all strong operations (when a strong operation *op* returns a stable response before another strong operations *op'* starts, then *op* will be serialized before *op'*).

Intuitively, *S* corresponds to the order of operations on the *committed* list. Note that in Creek the execution of operations always respects the order in which each replica invokes operations: weak operations invoked by each replica $p_i$ are ordered using their *timestamp*s and *id*s, both of which are monotonically increasing. When $p_i$ invokes a strong operation *op*, *op* has a timestamp that is at least as large as the *timestamp* of the last weak operation invoked by $p_i$. Hence the tentative execution of *op* will happen after the execution of all weak operations invoked previously by $p_i$. Moreover, *op*'s *id* is strictly larger than *id* of every weak operation invoked previously by $p_i$. Therefore, when the identifier of *op* is CAB-delivered and thus *op* moved from the *tentative* to the *committed* list, the relative order between *op* and weak operations previously invoked by $p_i$ is preserved (see the commit function).

---
[5]Creek does not make any assumptions on the semantics of operations issued to replicas other than that operations must be deterministic.

The real-time order among strong operations is satisfied in a straightforward way. Assume there is a strong operation *op*, such that some replica $p_i$ already returned a stable response of *op* to the client. Clearly, in order to produce a stable response, earlier $p_i$ must have had CAB-delivered an appropriate message concerning *op*. Assume that now some replica $p_j$ invokes some other strong operation *op'* and thus CAB-casts the identifier of *op'*. Naturally, $p_i$ can only CAB-deliver the identifier of *op'* after it had already CAB-delivered the identifier of *op*. By the guarantees of CAB, all replicas CAB-deliver all messages in the same order. Hence, similarly to $p_i$, all replicas CAB-deliver the message concerning *op'* after CAB-delivering the message concerning *op*. Thus on every replica *op* will appear on the replica's *committed* list before *op'*.

Creek satisfies *fluctuating eventual consistency (FEC)* for weak operations. Intuitively, FEC requires that each operation *observes* some serialization of (a subset) of operations already submitted to the system (each operation executes on a replica state that has been obtained by a serial execution of the operations given). The observed serialization corresponds to the *committed* · *tentative* list in Creek. When the identifiers of strong operations are CAB-delivered, operations are moved from the *tentative* to the end of the *committed* list. Hence, *committed* corresponds to the ever growing single serialization to which all serializations observed during executions of weak operations and tentative executions of strong operations gravitate. However, the serializations observed by different weak operations gravitate towards one, ever growing single serialization. In other words, any operation *op* can be observed only temporarily (a finite number of times) by other operations in an order which is not equal to the final execution order of *op*. ∎

## 4.4 High-performance protocol

**Read-only operations.** An obvious optimization of Creek involves executing weak read-only operations without performing any network communication with other replicas. However, this optimization does not address the core limitation of Creek, which comes from excessive number of operation rollbacks and reexecutions. To improve Creek's performance, we modified Creek in several ways. In our discussion below we focus on the updating operations.

**Rollbacks only if necessary.** Suppose there are two already executed operations $op_i, op_j \in$ *tentative*, and $op_i$ appears before $op_j$ on *tentative*. If $op_j$ is moved to *committed* (e.g., because $op_j$ is being committed and $op_i$ does not belong to the causal context of $op_j$), the basic version of Creek must rollback both operations and reexecute them but in the opposite order. However, if $op_i$ and $op_j$ operated on distinct data, no rollbacks or reexecutions are necessary (at least with respect to only these two operations). Typical workloads exhibit *locality*, i.e., the requests do not access all data items with uniform probability [48]. Hence, such an optimization brings dramatic improvement in Creek's performance.

**Multiversioning.** To facilitate efficient handling of situations similar to the one described above, we extended Creek with multiversioning scheme [15]. The modified version of Creek holds multiple immutable objects called *versions* for data items accessed by operations. Versions are maintained within a *version store*. Each version is created during execution of some operation *op* and is marked using a special timestamp that corresponds to the location of *op* on the *committed* · *tentative* list. The execution of any operation *op* happens in isolation, on a *valid* snapshot. It means that the snapshot includes all and only the versions created as the result of execution of all operations *op'*, such that *op'* appear before *op* on *committed* · *tentative* at the time of execution of *op*. A rollback of *op* does not remove the versions created during the execution of *op*. Instead, all versions created during execution of *op* are marked, so they are not included in the snapshots used during execution of all operations *op'* that start execution after the rollback of *op*.

A rollback of *op* may cascade into rollbacks of other operations. Suppose as before that there are two already executed operations $op_i, op_j \in$ *tentative* and $op_i$ appears before $op_j$ on *tentative*. Suppose also that $op_x$ is RB-delivered, and $op_x$ has a lower timestamp than $op_i$. In the basic version of Creek, both $op_i$ and $op_j$ would be rolled back and reexecuted after the execution of $op_x$. Thanks to multiversioning, we can execute $op_x$ on a consistent snapshot corresponding to the desired order of $op_x$ on *tentative* and then check, whether the execution of $op_x$ created new versions for any objects read by $op_i$. If not, we do not need to roll $op_i$ back and we can proceed to check in a similar way the possible conflict between $op_x$ and $op_j$. On the other hand, if $op_i$ is rolled back, we need to check for conflicts between $op_x$ and $op_j$ as well as between $op_i$ and $op_j$, because $op_j$ might have read some no longer valid versions created by $op_i$.

Note that one needs to be careful in garbage collecting versions. Since a newly RB-delivered operation can be placed in the middle of the *tentative* list, we need to maintain all versions produced during execution of the operations on the *tentative* list. We track live operations (operations being executed) to see which snapshots they operate on. This way we never garbage collect versions that might be used by live operations. Having that in mind, for each data item we can attempt to garbage collect all versions which were created during executions of operations $op \in$ *committed*, except for the most recently created value. We can also eventually remove all versions created by operations that were later rolled back (by specification of Creek, new transactions that start execution after the rollback already happened will not include the rolled back versions in their snapshots).

Under normal operation, when strong operations are committed every once in a while, the number of versions for each data item should remain roughly the same. However, when no strong operations are being committed (because no strong operations are submitted for a longer period of time or no message can be CAB-delivered due to a network partition), the number of versions starts to grow. We could counter such a situation by, e.g., periodically issuing strong no-op operations, that would gradually stabilize weak operations. Otherwise, we need to maintain all versions created by operations $op \in tentative$. In such case, we could limit the amount of data we need to store, by collecting complete snapshots (that represent some prefix of $committed \cdot tentative$), and recreate some versions when needed, by reexecuting some previously executed operations on the snapshot.

**Concurrent execution.** Multiversioning allowed us to relatively easily further extend Creek to support concurrent execution of multiple operations. Having multiple operations execute concurrently does not violate correctness, because each operation executes in isolation and on a consistent snapshot. The newly created versions are added to our version store after the operation completes execution. We do so atomically and only after we checked for conflicts with other concurrently executed operations which already completed their execution. In case of a conflict, we discard versions created during the execution and reexecute the operation.

### 4.5 Future work

Now we outline how Creek can be further extended to support partial replication. Doing so is far from straightforward and thus warrants a new, complex scheme. Hence we leave the detailed description of the partially replicated version of Creek for another paper.

We consider a system with $m$ disjoint data partitions (or shards). Operations submitted to the system are either *single partition operations (SPO)* or *multiple partition operations (MPO)*; we assume that the set of relevant shards for each operation is known *a priori*. SPOs can be either weak or strong operations, but MPOs must be strong operations. As in the basic variant of Creek, the contents of any operation are disseminated among replicas using a gossip protocol. Naturally, the messages are only sent to replicas that keep data relevant for the particular operation. Additionally, for any strong operation (and so all MPOs), the identifier of the operation is multicast to the relevant replicas using *conditional atomic multicast (CAM)*, a protocol based on atomic multicast [42] which is extended in a similar way to which CAB extends atomic broadcast. By using CAM to send only the operations' identifiers, CAM does not become the bottleneck.

The execution of SPO happens mostly as in the basic variant of Creek (see also below). On the other hand, the execution of MPOs proceeds similarly to cross-partition commands in S-SMR [33]: all replicas to which an MPO is issued execute the MPO independently, and eagerly exchange the values of read objects because they will be needed by other replicas that do not replicate the objects locally. More precisely, all write operations are buffered until the operation completes. When an operation reads an object replicated locally, the read value is multicast (using reliable multicast, not CAM) to the relevant replicas so they can also process the read. A replica that reads an object not replicated locally pauses the processing of the operation until the replica receives an appropriate value. If replica $p_i$ learns that it needs to reexecute an MPO (because, e.g., $p_i$ received some other operation with lower timestamp), $p_i$ informs other replicas about the event, reexecutes the MPO on a different snapshot, sends to other relevant replicas the new data they require to complete the reexecution of MPO, and waits for values it needs to perform the reads. Upon speculative execution, a tentative response is returned to client as in Creek. However, in order to move an MPO from the *tentative* to the *committed* list (and return a stable response to the client), a replica that executed the MPO must perform one additional check, as we now explain.

Assume that replica $p_i$ firstly speculatively executed MPO $op$ and then delivers through CAM the identifier of some strong SPO $op'$. In the final operation execution order $op'$ will be serialized before $op$ because $p_i$ has not yet delivered through CAM the identifier of $op$. The arrival of the identifier of $op'$ renders the speculative execution of $op$ invalid (if $op$ reads any objects modified by $op'$). However, the replicas that also execute $op$ but belong to other partitions than $p_i$ will not receive the information regarding $op'$ ($op'$ is an SPO and so reaches only replicas in $p_i$'s shard). Hence, unless $p_i$ notifies them, they could incorrectly move $op$ to their *committed* list. Thus, before any replica moves an MPO from the *tentative* to the *committed* list, it waits for a confirmation from one replica from each shard to which the MPO was issued. A replica multicasts (using reliable multicast, not CAM) a confirmation as soon as it delivers through CAM the identifier of $op$ (so it knows no other strong operation $op'$ can be serialized before $op$ on the *committed* list). Note that the above scenario is also possible for any two concurrent MPOs but sent to two different (but overlapping) sets of shards.

## 5 Implementing CAB

There is a strong analogy between CAB and atomic broadcast (AB) built using indirect consensus [14]. Our approach is quite a straightforward extension of the AB reduction to indirect consensus presented there, as we now discuss.

As shown in [49], AB can be reduced to a series of instances of distributed consensus. In each instance processes reach agreement on a set of messages to be delivered. Once the agreement is reached, messages included in the decision value are delivered in some deterministic order by each process. Indirect consensus reduces the latency in reaching agreement among the processes by distributing the messages (values being proposed by the processes) using a gossip protocol and having processes to agree only on the identifiers of the messages. Hence, a proposal in indirect consensus is a pair of values $(v, rcv)$, where $v$ is a set of message identifiers (and $msgs(v)$ are the messages whose identifiers are in $v$), and $rcv$ is a function, such that $rcv(v)$ is true only if the process has received $msgs(v)$. Indirect consensus' primitives are similar to the ones of classic distributed consensus: $propose(k, v, rcv)$ and $decide(k, v)$, where $k$ is the number identifying a concrete consensus execution. Naturally, whenever a decision is taken on $v$, indirect consensus must ensure that all correct processes eventually receive $msgs(v)$. We formalize this requirement by assuming eventual stable evaluation of $rcv(v)$.[6] Formally, indirect consensus requires:

- *termination*: if eventual stable evaluation holds, then every correct process eventually decides some value,
- *uniform validity*: if a process decides $v$, then $(v, rcv)$ was proposed by some process,
- *uniform integrity*: every process decides at most once,
- *uniform agreement*: no two processes (correct or not) decide a different value,
- *no loss*: if a process decides $v$ at time $t$, then for one correct process $rcv(v) = true$ at time $t$.

In indirect consensus, the $rcv(v)$ function explicitly concerns local delivery of messages, whose identifiers are in $v$. However, $rcv$ could be replaced by any function $f$ that has the same properties as $rcv$, i.e., eventual stable evaluation holds for $f$. In CAB, instead of $rcv(v)$, we use a conjunction of $rcv(v)$ and test predicates $q(m)$ for each CAB-cast message $m$, whose identifier is in $v$. This way we easily obtain an efficient implementation of CAB, because we minimize the sizes of propositions, on which consensus is executed. In practice, a very efficient implementation of CAB can be obtained by slightly modifying the indirect variant of Multi-Paxos [50]. Due to space constraints, we omit the more formal description of the complete reduction of CAB to indirect consensus.

## 6 Experimental evaluation

Since Creek has been designed with low latency in mind, we are primarily interested in the *on-replica latencies* (or simply *latencies*) exhibited by Creek when handling client operations (the time between a replica receives an operation from a client and sends back a response; the network delay in communication with the client is not included). From a client's perspective, important are the *stable latencies* for strong operations and the *tentative latencies* for weak operations: when an operation is marked as strong, it means it is essential for the client to obtain a response that is *correct* (i.e., it results from a state that is agreed upon by replicas). On the other hand, weak operations are to be executed in an eventually consistent fashion, so the client expects that the tentative responses might not be 100% accurate (i.e., the same as produced once the operations stabilize).

---

[6]In the original paper [49], this requirement has been called *hypothesis A*.

We compare the latencies exhibited by Creek with the latencies produced by other replication schemes that enable arbitrary semantics. To this end, we test Creek against SMR [21] [22], Archie [23] (a state-of-the-art speculative SMR), and Bayou [20] (mainly due its historical significance). For all systems we also measure the average CPU load and network congestion. Moreover, for Creek and Archie we check the accuracy of the speculative execution, i.e., the percentage of weak operations, for which the first speculative execution yielded results that match the ultimate results corresponding to this operation. Archie, as specified in [23], does not return tentative responses after completing speculative execution. We can however predict what would be the tentative latency for Archie and thus we plot it alongside stable latency.

Recall that Creek (similarly to Bayou) uses a gossip protocol to disseminate (both weak and strong) operations among replicas. To ensure minimal communication footprint of the inter-replica synchronization necessary for strong operations, Creek uses an indirect consensus-based implementation of CAB. On the other hand, Archie and efficient SMR implementations (e.g., [51]) disseminate entire messages (operations) solely through atomic broadcast (AB). Since our goal is to conduct a fair comparison between the schemes, our implementations of SMR and Archie rely on a variant of AB that is also based on indirect consensus.

### 6.1 Test environment

We test the systems in a simulated environment, which allows us to conduct a fair comparison: all systems share the same implementation of the data store abstraction and the networking stack, and the sizes of the exchanged messages are uniform across systems (apart from additional messages exchanged through CAB in Creek). We simulate a 5-replica system connected via 1Gbps network. Each replica can run up to 16 tasks in parallel (thus simulating a 16-core CPU). The network communication latencies are set to represent the typical latencies achieved in a data center (0.2-0.3 ms).

For our tests we use TPC-C [19], a popular OLTP benchmark that simulates a database system used by a wholesale parts supplier operating a number of warehouses (the number of warehouses functions as a scale factor for the benchmark; in our tests it is set to 5, what translates into medium contention levels observed between concurrently executed transactions). TPC-C defines 5 types of transactions, which are chosen at random and then submitted to the system (in brackets we give the probability with which the transaction is being chosen): *New order* (45%), *Payment* (43%), *Delivery* (4%), *Order status* (4%), *Stock level* (4%). The most *sensitive* of the transactions is the *Payment* transaction, which, among others, updates the customer's balance. Hence, in Creek's tests we mark all *Payment* transactions as strong operations. All other transactions are run as weak operations following the *act now, apologize later* principle (note that the *Order status* and *Stock level* transactions are read-only).

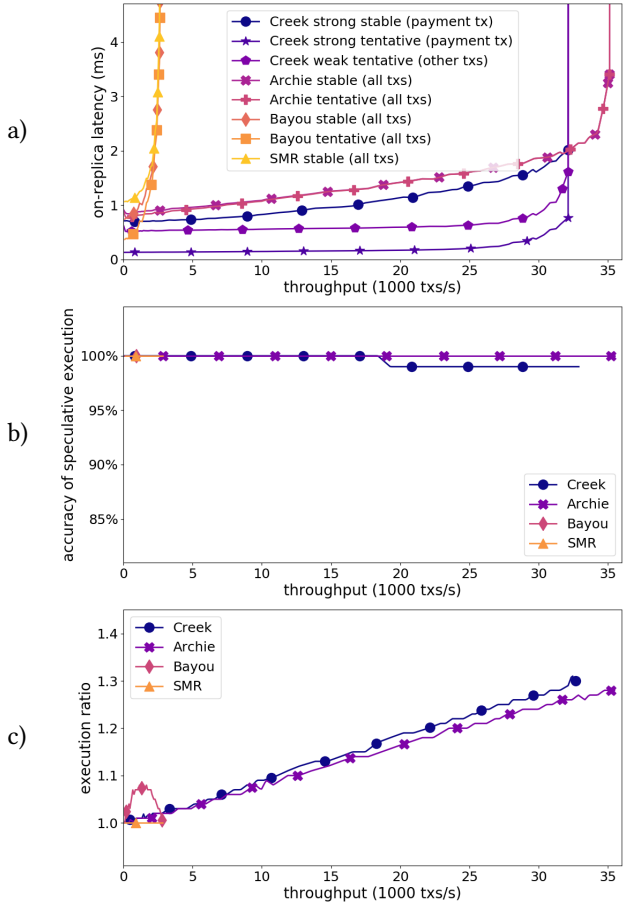

**Figure 1.** Test results for scenarios with medium contention (TPC-C is setup with scale factor 5).

### 6.2 Test results

In Figure 1a we present the on-replica latencies for all systems in the function of achieved throughput. In all tests the network is not saturated for any of the systems: messages exchanged between the replicas are small and transactions take a significant time to execute.

SMR and Bayou, whose maximum throughput is about 2.7k txs/s, are easily outperformed by Creek and Archie, both of which take advantage of multicore architecture. The latter systems' peak throughputs are about 32.5k and 35k txs/s, respectively. Since in our tests the network is not saturated, for all systems the peak throughput is when the CPU is fully utilized, i.e., when the backlog of unprocessed transactions starts to build up (as signified by the latency striking up).

When CPU is not saturated, Creek's latency for tentative responses directly corresponds to the transaction execution times (no network communication is necessary to return a tentative response): for transactions executed as weak operations is steady around 0.5 ms, whereas for the *Payment* transaction is about 0.1ms (the *Payment* transactions are relatively short; recall that for these transactions we are focused on the stable response latency). Creek's latency in obtaining a stable response for a *Payment* transaction is a few times higher (0.8-1.2 ms), because producing the response involves inter-replica synchronization, i.e., an operation's identifier needs to be CAB-cast. Therefore, under our assumptions and using a Paxos-based implementation of CAB, the operation identifier can be CAB-delivered after 3 communication phases, which adds at least about 0.7-0.8 ms to the latency.

Both Creek's latencies for weak tentative responses and for stable strong responses are significantly smaller compared to the latencies achieved by Archie (Creek's latencies are 37-61% and 75-85% of those exhibited by Archie, respectively). It is because before an Archie's replica can start processing a transaction, it first needs to broadcast and deliver it. More precisely, an Archie replica starts processing a transaction upon optimistic delivery of a message containing the transaction, which was sent using AB. The speculative execution in Archie has little impact on stable latency: on average, before a speculative transaction execution is completed by an Archie replica, an appropriate AB message is delivered by the replica, thus confirming the optimistic delivery order (hence the perfect accuracy of the speculative execution, see Figure 1b). It means that a replica starts the execution of a transaction a bit earlier than it takes for a message to be delivered by AB, thus yielding 5-8% reduction in latency (observable only for the low throughput values).[7]

Returning tentative responses makes little sense, when most of the time they are incorrect (they do not match the stable responses). Our tests show, however, that the tentative responses produced by Creek are in the vast majority of cases correct: the accuracy of the speculative execution ranges between 98-100% (see Figure 1b).

As evidenced by Figure 1c, the *execution ratio* (the average number of executions performed of each transaction submitted to the system) is slightly higher for Creek compared to Archie's due to Creek's higher variance in the relative order between tentatively executed transactions. This fact explains the slightly lower maximum throughput of Creek when compared to Archie's. Archie curbs the number of rollbacks and reexecutions by allowing the replicas to limit the number of concurrently executed transactions. Moreover, in Archie, when the leader process of the underlying AB does not change, the optimistic message delivery order always matches the final delivery order.

---

[7] The impressive speed-up achieved by Archie, as described in the original paper [23], was due to network communication latencies, which were about 3-4ms, over 10 times higher than the ones we assume in our tests.

SMR executes all transactions sequentially, after they have been delivered by AB. It means that SMR exhibits high latency compared to Creek and Archie, and has very limited maximum throughput. Bayou cuts the latency compared to SMR, because Bayou speculatively executes transactions before the final transaction execution order is established. However, its maximum throughput is comparable to SMR's, as Bayou also processes all transactions sequentially.

### 6.3 Varying test parameters

Recall that in TPC-C, by changing the number of warehouses in the benchmark configuration, we can change the contention level: the greater the number of warehouses, the smaller the contention. We conducted additional tests with the number of warehouses set to 1 (the high contention scenario) and 20 (the low contention scenario). In terms of the observed latency, the relative differences between the systems were similar to the ones we discussed in the context of the main results (hence we omit the plots with detailed results). The change in the contention level impacted the overall performance: in the low contention scenario, both Creek and Archie reached about 40k txs/s, and in the high contention scenario, these systems achieved about 11k txs/s (the performance of SMR and Bayou stayed roughly the same as before, because these systems are essentially single threaded). Naturally, higher contention levels negatively impacted the execution ratio, which increased for both Creek and Archie to about 1.8. The accuracy of speculative execution in Creek was slightly lower than before and ranged between 92% and 100%. On the other hand, when contention was low, the execution ratio never exceeded 1.1 for both Creek and Archie, and both systems always achieved perfect accuracy of speculative execution.

Increasing the fraction of strong transactions in the workload means that the latency of stable responses for strong transactions in Creek is a bit closer to Archie's latency. It is because now, on average there are fewer transactions in the causal context of each strong transaction, and thus the transaction can be CAB-delivered earlier. The smaller causal contexts also translate into a slightly higher execution ratio, as fewer transactions can be committed together (recall that a strong transaction *stabilizes* weak transactions from its causal context upon commit). Increasing the fraction of strong operations in the workload naturally did not impact neither the performance of SMR nor Bayou.

Now we consider what happens when transactions take longer to execute. In the additional tests we increased the transaction execution time five times. Understandably the maximum throughput of all systems decreased five times. The maximum execution ratio for both Creek and Archie were lower than before, because there were fewer transactions issued concurrently. Longer execution times also meant that the inter-replica communication latency had smaller influence on the overall latency in producing (stable) responses (execution time dominates network communication time). In effect, when the fraction of strong operations is high (50%), the stable latency in Creek matches the (tentative/stable) execution latency in Archie, and the latency of Bayou is closer to SMR's. When the fraction of strong operations is relatively low (10%), the latency for Creek is lower compared to Archie's due to the same reasons as before.

Understandably, using machine clusters containing more replicas do not yield better performance, because all tested replication schemes assume full replication. Consequently every replica needs to process all operations submitted by the clients. To improve the horizontal scalability of Creek, it needs to be adapted to support partial replication (see Section 4.5). We leave that for future work. Using CPUs with more cores has no effect on SMR and Bayou, but allows Creek and Archie to (equally well) handle higher load.

### 6.4 Performance analysis

As shown by the TPC-C tests, Creek greatly improves the latency compared to Archie, the state-of-the-art speculative SMR system, and also provides much better overall throughput than SMR and Bayou.[8] In fact, the tentative latency exhibited by Creek is up to 2.5 times lower compared to Archie's. Moreover, even though strong operations constituted over 40% of all operations in the workload, Creek improves the stable latency by 20% compared to Archie (when the percentage of strong transaction is high, these latencies exhibited by Creek and Archie are comparable). Crucially, the tentative responses provided by Creek for both weak and strong transactions are correct in the vast majority of cases.

Similarly to Bayou, but unlike Archie and SMR, Creek remains available under network partitions (naturally, stable responses for strong transactions are provided only in the majority partition, if such exists). Under a heavy workload, Creek might take a long time to reconcile system partitions once the connection between the partitions is reestablished: the execution order of many transactions needs to be revisited, and some of them might be reexecuted. However, if operations issued by clients connected to replicas in one partition do not conflict with the operations issued by clients to replicas in some other partition, no unnecessary operation reexecutions are needed. Note that if we made some additional assumptions about the operation semantics, in some cases we could facilitate efficient merging of replica states from different partitions, as in CRDTs [8].

Naturally, eventually consistent systems which disallow unrestricted mixing of strong and weak operations and restrict the semantics of operations (e.g., to CRUD), such as Apache Cassandra [3], can perform much better than Creek. It is because these systems limit or avoid altogether operation reexecutions resulting from changes in the order in which the updates are processed. However, as we argued in Section 1, these systems are not suitable for all applications and they are difficult to use correctly.

---

[8]The improvement would be even more pronounced if we had not used an optimized version of atomic broadcast for Creek's competitors.

## 7 Conclusions

In this paper we presented Creek, a proof-of-concept, eventually-consistent, replication scheme that also enables execution of strongly consistent operations. By its design, Creek provides low latency in handling operations submitted by the clients and yields throughput that is comparable with a state-of-the-art speculative SMR scheme. Creek does so while remaining general: it supports execution of arbitrary complex, deterministic operations (e.g., SQL transactions). We believe that the Creek's principle of operation can be used as a good starting point for other mixed-consistency replication schemes which are optimized for more specific use.